\documentclass[twocolumn,english,showpacs,preprintnumbers,amsmath,amssymb,aps,prb]{revtex4}
\usepackage[T1]{fontenc}
\usepackage[latin9]{inputenc}
\usepackage{graphicx}
\usepackage{esint}
\usepackage{epstopdf}

\makeatletter

\@ifundefined{textcolor}{}
{%
 \definecolor{BLACK}{gray}{0}
 \definecolor{WHITE}{gray}{1}
 \definecolor{RED}{rgb}{1,0,0}
 \definecolor{GREEN}{rgb}{0,1,0}
 \definecolor{BLUE}{rgb}{0,0,1}
 \definecolor{CYAN}{cmyk}{1,0,0,0}
 \definecolor{MAGENTA}{cmyk}{0,1,0,0}
 \definecolor{YELLOW}{cmyk}{0,0,1,0}
 }

\makeatother

\usepackage{babel}
\begin{document}

\title{Random Organization in Periodically Driven Gliding Dislocations }

\author{C. Zhou$^{1,2}$, C. J. Olson Reichhardt$^{2}$, C. Reichhardt$^{2}$,
and I. Beyerlein$^{2}$}

\affiliation{$^{1}$Department of Materials Science \& Engineering,
Missouri University of Science and Technology, Rolla, Missouri 65409, USA\\
$^{2}$Center for Nonlinear Studies and Theoretical Division, Los Alamos National Laboratory, Los Alamos, New Mexico 87545, USA }

\date{\today}
\begin{abstract}
We numerically examine 
dynamical irreversible to reversible transitions and random organization for 
periodically driven gliding dislocation assemblies using the 
stroboscopic protocol developed   
to identify random organization in periodically driven 
dilute colloidal suspensions. 
We find that the gliding dislocations 
exhibit features associated with random organization
and evolve into  
a 
dynamically reversible state after a transient 
time extending over a number of cycles. 
At a critical 
shearing amplitude,
the
transient time diverges. 
When the dislocations enter the reversible state they organize into patterns
with fragmented domain wall type features.  
\end{abstract}


%keywords:
%Dislocations, Nonequilibrium phase transitions, Pattern formation, 
%Reversibility

\maketitle
\section{Introduction}
Recent experiments and simulations on 
non-thermal dilute colloidal systems under periodic external driving 
\cite{19,1,18,3,16}  
have revealed a transition from irreversible to reversible
behavior in the colloidal particle positions after 
a number of cycles of the external 
drive are applied \cite{19,18,1,12}. 
The different regimes are identified by marking
the particle positions 
at the beginning of each drive cycle 
and comparing them to the positions at the end of the cycle \cite{19,1}.
In general the system always behaves
irreversibly for the first few cycles, 
but either settles into a reversible state or        
remains irreversible. 
As described in Ref.~\cite{19},
in the irreversible regime, the particle positions differ at the beginning
and end of the cycle, 
and the accumulated displacements over many cycles 
correspond to an anisotropic random walk. 
In the reversible regime, the particles all return to the  same positions
at the end of each cycle so there is no diffusive behavior or particle mixing.  
For low driving amplitudes $\tau_{ext}$, 
the system evolves into a reversible state 
after a transient time $t_{n}$ of only a few cycles.   
As 
$\tau_{ext}$ increases, $t_{n}$ increases and
undergoes a power law divergence at a critical driving amplitude 
$\tau_c$. 
For drives above $\tau_c$,
the system remains 
irreversible for an arbitrary number of cycles, but there
is still a transient decay into the steady irreversible state during
a time $t_n$, which 
diverges as a power law near $\tau_c$. 
The dilute colloidal particles rearrange themselves
in the reversible state 
so that they no longer interact, and since
these states 
lack spatial organization, they were termed
'randomly organized.' 
The power law divergence of 
$t_n$ 
has been interpreted\cite{20} to indicate that 
the reversible to irreversible transition 
is a nonequilibrium phase transition. 

Reversible to irreversible transitions and random organization 
in periodically driven systems with
quenched disorder have also been studied for 
vortices in type-II superconductors, where behavior very similar to
that of the dilute colloidal suspensions occurs:  
the system organizes into reversible states and has divergent 
transient times on either side of a critical drive amplitude \cite{2,6,10,17}. 
The vortices
have much longer range 
interactions
than the colloidal particles, and in the
reversible state 
the vortices are still strongly interacting with each other, indicating
that the idea of random organization can be extended 
beyond systems with simple contract interactions. 
Random organization has also been studied in the context of plastic 
depinning of colloidal particles and vortices, 
where a divergent transient time appears at the 
pinned to sliding transition \cite{14,9,8,21}.   
An intriguing question is whether there are other systems 
with long range interactions that can also exhibit a 
dynamical reversible to irreversible transition under 
periodic driving, and 
if there can be reversible states that are not simply random 
but form patterns or structures that differ from the initial 
random distributions. 
Dislocations undergoing glide in materials are a prime candidate system
in which to address these questions.
Under dc loading, the stress in
materials containing dislocations 
increases with strain and levels off when yielding occurs and
the material flows \cite{25,27}.
A true elastic regime usually occurs only for very low strains,
while in the regime
where the stress is increasing nonlinearly, plastic rearrangements occur.
In general, plastic flow
is thought of as always being irreversible; however, 
there are now examples of reversible flow of dislocations    
in what is called recovery when the system is driven in a 
single cycle \cite{30,31}. 
Additionally, recent simulations of amorphous solids under periodic
forcing have shown that just as in the dilute colloidal case, the system can
begin in an irreversible plastic
state and then settle into a reversible state with plastic rearrangements
that repeat periodically, and that there is a critical amplitude
above which the motion remains irreversible 
\cite{I,S}.
Experiments on periodically sheared jammed materials have also identified
a reversible to irreversible transition \cite{A}, and numerical simulations
have explored the connection between jamming and reversible-irreversible
transitions \cite{C}.

Here we use large scale numerical simulations to examine 
the periodic forcing of gliding dislocations
for different shear amplitudes with the stroboscopic 
measure developed for the dilute colloidal system.
Our simulations are based on a well-established dislocation glide model that 
has previously been employed to study
intermittency \cite{23,28,29}, avalanches \cite{26}, creep \cite{24}, hysteresis \cite{22}, jamming \cite{27}, and driven phases \cite{ourstuff}.     
By comparing the dislocation positions from one drive cycle to the next, 
we can identify
the number $t_n$ of cycles required for the system to 
organize into a state where the
dislocations return to the same position after each cycle. 
For low shear amplitude, the system quickly  settles into a
reversible state. 
For large loading,
the dislocations pass each other so rapidly that they interact only
weakly,
and the system again
quickly organizes to a reversible state. At a critical load
$\tau_c$,  we find that $t_n$ grows rapidly,
similar to what is observed  
in the colloidal and vortex systems, and that $t_n$ can be 
fitted to a power law with exponents consistent with
those obtained in other systems \cite{1,6,9,14}.  
Unlike the previously studied systems, our dislocation model
always organizes to a reversible state both above and below
$\tau_c$, with 
a partially random reversible state appearing in the low drive
regime, 
and a state with more well-defined wall structures occurring in the
high drive regime.  

\section{Computational Details}
We model dislocations in a two-dimensional (2D) system 
that are each confined to glide on parallel slip planes \cite{23,24,27}.  
The motion of dislocation $i$ depends on its Burgers vector, 
dislocation-dislocation
interactions, and the external load 
according to
the following equation of motion: 
\begin{equation}
\eta\frac{dx_{i}}{dt} = b_{i}\left(\sum^{N}_{j\neq i}\tau_{int}
({\bf r}_{j} - {\bf r}_{i}) - \tau_{\rm ext}\right) .
\end{equation}  
Here $\eta$ is an effective damping constant that 
arises from dissipation mechanisms such as the radiation of phonons by
dislocations, and
the Burgers vector value 
$b_{i}$ 
can be either positive or negative.   
The dislocations interact via an anisotropic stress field  
given by $\tau_{int}(r) =  b\mu[x(x^{2} - y^{2})/(2\pi(1-\nu)(x^{2} + y^{2})^2]$ where
${\bf r} = (x,y)$,  $\mu$  
is the shear modulus, and $\nu$ is the Poisson ratio. 
Here $|b| = 1$, $\eta = 1.0$,
and $\mu/2\pi(1-\nu) = 1.0$.   
The external load is $\tau_{\rm ext}$ and the resulting dislocation motion
is in either the positive or negative $x$-direction 
depending on the sign of the Burgers vector. 
We impose the rule that
dislocations must be separated by a vertical
distance no less than $\delta y$ in order to avoid
dislocation annihilation processes, and place only one dislocation
on each glide plane.
We measure the absolute value of the dislocation velocity 
$v$ versus the external 
load. 
In previous work with this system, yielding and intermittent dynamics
were studied under dc loads \cite{24,27}.
Here we consider the effects of periodic loading and 
compare the positions of the dislocations at the beginning and end of
each load cycle.
The external load 
$ \tau_{\rm ext}(t) =  \tau_{\rm ext}{\rm sgn}(\sin(2\pi t/P))$ is a 
square 
wave with total amplitude $2\tau_{\rm ext}$ and period $P$ measured in
simulation time steps.
Unless otherwise noted, we take $P=2.5 \times 10^5$.
The fraction $A_n$ of dislocations  
that do not return to the same
position are labeled as active and identified 
by comparing the net dislocation displacement after $n$ cycles
with 
a small distance $\delta x$,
\begin{equation}
A_{n} = N_{d}^{-1}\sum^{N_{d}}_{i= 1}\Theta\left(\left[X_{i}(t_{n+1} ) - X_i(t_{n})\right]^2 - \delta x\right)     ,
\end{equation}
where $\Theta$ is the Heaviside step function.
We take $\delta x=10^{-5}$; we have tried 
various other cutoffs and find that 
smaller values of $\delta x$ give the same results.
When the system is in a reversible regime, $A_n = 0.0$.  
We consider systems containing up to 480 dislocations 
with equal numbers of positive and negative Burgers vectors. 
The dislocations are allowed to relax for a fixed time before the external driving is applied. 
In general the system always shows an initial irreversible dynamics 
during the first few cycles,
as also observed in the colloidal suspensions \cite{1}  
and vortex experiments \cite{6}.

\begin{figure}
\includegraphics[width=3.5in]{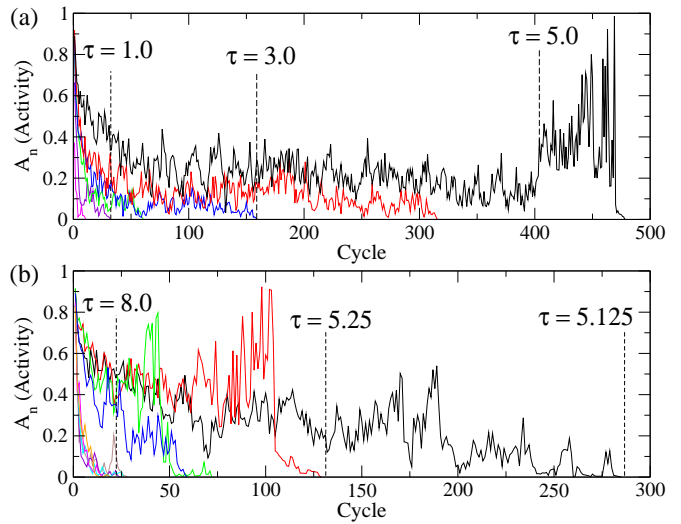}
\caption{ 
(a) The fraction of active dislocations $A_{n}$ that do 
not return to the same position after 
$n$ successive cycles of a periodic drive, plotted vs $n$. 
When $A_{n}= 0$, the system is considered reversible. 
From left to right, 
the external drive $\tau_{\rm ext} = 0.5$, 1.0, 2.0, 3.0, 4.0, and $5.0$.
The $\tau_{\rm ext} = 1.0$, 
3.0 and $5.0$ curves are highlighted to show that the transient times 
increase with increasing $\tau_{\rm ext}$.
(b) The same for $\tau_{\rm ext} = 5.125$, 5.25, 5.5, 6.0, 8.0, and $9.0$, 
with the $\tau_{ext} = 5.125$, 5.25, and $8.0$ curves marked
to show that the transient times are now decreasing
with increasing $\tau_{\rm ext}$.     
}
\label{fig:1} 
\end{figure}

\begin{figure}
\includegraphics[width=3.5in]{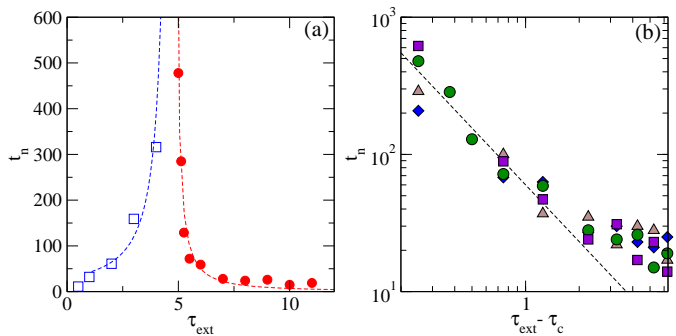}
\caption{ 
(a) $t_n$, the number of cycles required to reach a 
reversible state, vs drive amplitude $\tau_{\rm ext}$. 
Here $t_n$ passes through a maximum at finite $\tau_{\rm ext}$. 
The dashed lines are fits to
$|\tau_{c} - \tau_{\rm ext}|^\beta$ with $\tau_{c} = 4.75$ and $\beta = 1.375$.   
(b) A log-log plot of 
$t_n$ vs drive amplitude $\tau_{\rm ext}-\tau_c$
for the regime where $\tau > \tau_{c}$.
The drive period is $P=3.0 \times 10^5$ (squares), 
$2.5 \times 10^5$ (circles), $2.0 \times 10^5$, (triangles) 
and $1.5 \times 10^5$ (diamonds). 
The dashed line is a power law fit with
$\beta = 1.375$. 
}
\label{fig:2}
\end{figure}

\section{Results and Discussion}
\subsection{Transient times}
Under a small external dc 
strain, 
the velocities settle to zero so that 
the material can be said to be jammed or below yield. 
As the dc external drive is increased,
there is a critical yield load at which the velocity 
always remains finite; above this load, the system 
is beyond the yield point \cite{32}. 
From dc measurements with a slowly increasing external load, 
a well defined yield point $\tau_c$ can be obtained 
that 
increases as either $1/\sqrt(\rho)$ \cite{27}  or $1/\rho$ 
for increasing dislocation density $\rho$. 
In Fig.~\ref{fig:1}(a) we plot $A_{n}$ vs $n$          
for a system with a fixed frequency periodic drive for 
different values of $\tau_{\rm ext} = 0.5$
to $5.0$.
We highlight the transient times $t_n$ for drives  
$\tau_{\rm ext}=1.0$, 3.0, and $5.0$ to show that 
$t_n$ 
increases with increasing $\tau_{\rm ext}$.
Here, $A_n$ for $\tau_{\rm ext} = 0.5$ drops to zero after only 
$n=10$ cycles,
while for $\tau_{\rm ext} = 5.0$, 
reversibility   
does not appear until after $n=480$ cycles. 
Figure \ref{fig:1}(b) shows $A_n$ for $\tau_{\rm ext} = 5.125$, 
to $9.0$ with      
the drives at $\tau_{\rm ext}=5.125$, 5.25, 
and $8.0$ labeled to indicate that 
$t_n$ is now decreasing with increasing $\tau_{\rm ext}$. 
This result indicates that there is a peak 
in 
$t_n$ centered around a critical amplitude $\tau_{c}$. 

In Fig.~\ref{fig:2}(a) we plot the transient time $t_{n}$ 
vs $\tau_{\rm ext}$ to more clearly show the divergence in $t_n$ 
just below $\tau_{\rm ext}=5.0$. 
In the colloidal work, the transient times 
also showed a divergence centered at a critical drive amplitude that
was fit to a power law form \cite{1}. 
In Fig.~\ref{fig:2}(a), the lines are fits
to $t_n \propto |\tau_{\rm ext} - \tau_{c}|^{-\beta}$ 
with $\beta = 1.375$ and $\tau_{c} = 4.75$.  The fit of the data is best
for $\tau_{\rm ext} > 4.75$. 
For the 2D dilute suspension simulations, 
similar fits gave $\beta = 1.33$ \cite{1}, 
while the 3D experiments gave $\beta = 1.1$ \cite{1}. 
Experiments on periodically sheared superconducting 
vortex systems using the same 
fitting produced $\beta = 1.3$ \cite{6}. 
For simulations
of dc depinning in the plastic regime,
a diverging time scale at a critical depinning force was found
with $\beta = 1.37$ \cite{14},  while
experiments on dc driven 
superconducting vortices gave $\beta = 1.6$ \cite{9} and $\beta = 1.4$ \cite{6}.
In Fig.~\ref{fig:2}(b) we plot log-log fits of $t_{n}$ vs $(\tau-\tau_{c})$
for varied drive periods $P=3.0 \times 10^5$, $2.5 \times 10^5$, 
$2.0 \times 10^5$, and $1.5 \times 10^5$
for $\tau_{ext} > \tau_{c}$ 
with $\tau_{c} = 4.75$. 
The dashed line is a fit to $\beta = 1.375$.  The fit
breaks down for large $\tau_{\rm ext}$, as expected 
since critical phenomena should
only occur near $\tau_{c}$. 
Although our results are not able to provide a 
highly accurate exponent, they show that our 
system exhibits the same general trends found 
for the random organization observed in other systems,
with the exponents in reasonable agreement. 
This suggests that all these systems exhibit nonequilibrium dynamical
phase transitions in the same universality class. 

\begin{figure}
\includegraphics[width=3.5in]{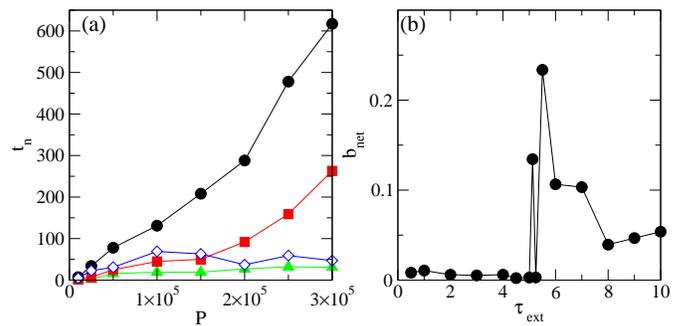}
\caption{
(a) 
$t_n$ vs cycle period $P$ 
for  $\tau_{\rm ext} = 8.0$ (diamonds), $5.0$ (circles),  $3.0$ (squares),
and $1.0$ (triangles). 
(b) The net burgers vector $b_{\rm net}$ 
along quasi-1D stripes vs $\tau_{\rm ext}$ 
for a system with $P=2.5\times 10^5$.
For $\tau_{\rm ext} > \tau_{c}$, the system starts to form domain walls.      
}
\label{fig:3}
\end{figure}

In Fig.~\ref{fig:3} we examine how the transient times $t_n$ change for a fixed 
periodic drive amplitude 
but with increasing drive period $P$. 
We find that the value of $\tau_{\rm ext}$  at which the divergence 
occurs is independent of $P$; however, near $\tau_c$ the
transient times increase with increasing $P$.
In Fig.~\ref{fig:3}(a) at $\tau_{ext} = 3.0$ and $\tau_{ext} = 8.0$, 
$t_n$ remains
constant for increasing $P$, 
while for  $\tau_{ext} = 5.0$ there is a strong increase in $t_{n}$ with
increasing $P$. 
At $\tau_{ext} = 3.0$ there is also a weaker increase in $t_n$ with 
increasing $P$. 
This result suggests that for very long
drive cycle period, the divergence in $t_n$ will become more pronounced.  

One clear difference between the dislocation system 
and the simulations of sheared dilute colloidal particles
is that the gliding dislocations
always evolve into a reversible state, whereas 
the colloidal system 
remains in a steady irreversible state 
at large drives. 
This may be a result of the fact that in
the dislocation model we employ, even though the interactions are 
2D, the dislocations are confined to move only along 1D lines \cite{23,24}.  
In the colloidal system, the particles can  move in the 2D continuum,
permitting the formation of many more possible states. 
In the 
superconducting vortex system, the vortices can freely move in the 2D plane and
there is again 
a transition to a steady irreversible 
state at high enough drive
\cite{2}.  
For more complicated dislocation dynamics models that
incorporate both climb and glide, 
it may be possible for the system to remain irreversible above the 
critical amplitude.  
It was proposed in the colloidal shearing work and the driven vortex systems
that the transition from the reversible to the irreversible state is a 
nonequilibrium phase transition or an absorbing phase
transition which may fall in the universality class of 
directed percolation or conserved directed percolation \cite{1,20}. 
The exponents that have been reported in these systems 
are consistent with this scenario, 
although it has not been possible to distinguish between the 
two classes since their exponents are nearly identical.
In our system, the divergence in 
$t_n$ 
suggests that a nonequilibrium phase transition occurs that is similar to that
found in the colloidal system, and that the dislocation system organizes
into two different absorbing states
on either side of the transition. 
In real experiments, additional effects such as climb may occur that could
change the nature of the transition.

\begin{figure}
\includegraphics[width=3.5in]{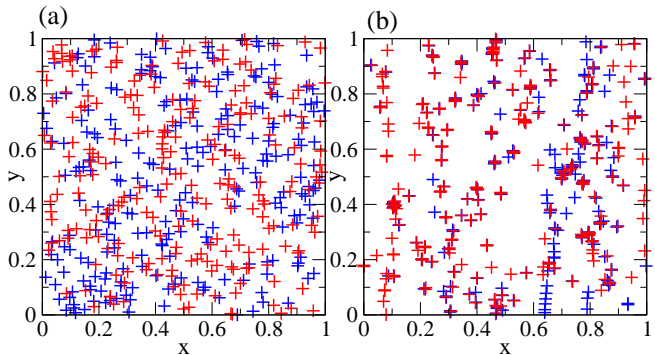}
\caption{
Dislocation positions, with the different shadings corresponding to 
the two different Burgers vectors.
(a) The undriven case showing a random 
spatial distribution of dislocations. 
(b) At $\tau_{ext} = 4.0$ and $P=2.5 \times 10^5$ time steps, 
the system forms a disordered configuration with a small number of wall-type
structures.
}
\label{fig:4}
\end{figure}

Although the high and low drive states in the dislocation system
cannot be distinguished by whether they are dynamically reversible or not, 
they do have different spatial dislocation structure properties. 
Figure \ref{fig:4}(a) shows that the dislocation positions without any 
external driving are strongly disordered. 
When the system reaches a reversible state for  
$\tau_{\rm ext} < \tau_{c}$, the dislocations still 
have a disordered spatial pattern; however, there are small 
regions where wall-like structures form, 
as illustrated in Fig.~\ref{fig:4}(b) 
for the reversible regime at $\tau_{\rm ext} = 4.0$. 
For $\tau_{\rm ext} > \tau_{c}$, the system forms a
larger number of  
wall structures.

To characterize the change in the dislocation structures across the 
critical drive, 
we measure the fraction of dislocations $P_{b}$ that are 
contained in walls by measuring 
the net Burgers vectors along thin $y$-direction strips of the system.
For random dislocation distributions in the regime  
$\tau_{\rm ext} < \tau_{c}$, $P_{b}$  is close to zero, while
for $\tau_{\rm ext} > \tau_{c}$, $P_{b}$ jumps up to a much higher value. 
This result shows that the dislocations do not necessarily organize into
a random state, 
as found for the colloidal systems \cite{1}; 
instead, for $\tau_{\rm ext} > \tau_c$ they can organize
into an absorbing pattern-forming state.
This suggests that the protocol of periodic forcing followed by a measurement
of the time required for the system to organize into a dynamically reversible
state might be fruitfully applied to
a much wider class of systems, including 
many pattern forming systems. 
For example, in 
driven twin boundaries, magnetic domain walls, 
and binary mixtures, one reversible 
state could be a random or disordered pattern 
while the other reversible state could be spatially ordered.

\section{Conclusion}
In summary, we have examined the dynamics of periodically 
driven gliding dislocations
in 2D simulations. We find that this system exhibits many of the same features
observed for random organization of periodically sheared colloidal particle
suspensions, granular media, and
superconducting vortices. The system is initially irreversible
with the dislocations
returning to different positions after each driving cycle, but
over time it organizes into a reversible state where the dislocations 
return to the same positions
after each drive cycle. 
The transient time required to reach a reversible state shows a 
divergence at a critical load near the dc yield point, and the exponents
of the divergence are consistent with those found for the 
colloidal assemblies, suggesting that the dislocation system may exhibit
a nonequilibrium phase transition of the same university 
class as the diluted colloidal systems. One difference from 
previous studies is that the dislocation system
always organizes into a reversible state,  
whereas the other systems remain in an irreversible state
above a critical driving amplitude.
The two reversible states into which the dislocations organize are
characterized by different patterns on
either side of the critical load.
For smaller loads, the system organizes into a more random 
state of small walls, while above the critical load,
the system organizes into states with partial wall structures. Future directions
would be to add more complicated dislocation dynamics such as      
cross slip and climb, which could change one of the reversible phases 
into an irreversible state that more closely resembles
that found in the periodically driven colloidal system.

\acknowledgments
This work was carried out under the auspices of the U.S.
DoE at LANL under Contract No. DE-AC52-06NA25396.

\end{document}